\documentclass[aps,twocolumn,showpacs,preprintnumbers,amsmath,amssymb,superscriptaddress]{revtex4}

\usepackage{graphicx}
\usepackage{dcolumn}
\usepackage{amsmath,amssymb}

\newcommand{\bm}[1]{\mbox{\boldmath{$#1$}}}

\def\cos{{\rm cos}}
\def\sin{{\rm sin}}

\makeatletter
\def\btt#1{\texttt{\@backslashchar#1}}%
\DeclareRobustCommand\bblash{\btt{\@backslashchar}}\makeatother

\begin{document}

\title[Short Title]{Topological Identification of Spin-1/2 Two-Leg Ladder with
Four-Spin Ring Exchange}
\author{I. Maruyama}%
\email{maru@pothos.t.u-tokyo.ac.jp}
\affiliation{%
Department of Applied Physics, The University of Tokyo, 7-3-1 Hongo,
Bunkyo-ku, Tokyo 113-8656, Japan }%
\author{T. Hirano}
\email{hirano@pothos.t.u-tokyo.ac.jp}
\affiliation{%
Department of Applied Physics, The University of Tokyo, 7-3-1 Hongo,
Bunkyo-ku, Tokyo 113-8656, Japan }%
\author{Y. Hatsugai}%
\email{hatsugai@sakura.cc.tsukuba.ac.jp}
\affiliation{%
Institute of Physics, 
University of Tsukuba, 1-1-1 Tennodai, Tsukuba, Ibaraki 205-8571, Japan }%

\date{\today}
\begin{abstract}
A spin-1/2 two-leg ladder with four-spin ring exchange 
is studied by quantized Berry phases,
used as local order parameters.
Reflecting local objects,
non-trivial ($\pi$) Berry phase is founded
on a rung for the rung-singlet phase
and on a plaquette for the vector-chiral phase.
Since the quantized Berry phase is
topological invariant
for gapped systems
with the time reversal symmetry,
topologically identical models
can be obtained
by the adiabatic modification.
The rung-singlet phase is
adiabatically connected to a decoupled rung-singlet model
and 
the vector-chiral phase is connected to 
a decoupled vector-chiral model.
Decoupled models reveals that the local objects are
a local singlet
and
a plaquette singlet
respectively.
\end{abstract}

\maketitle
\section{Introduction}
The recent progress in the multiple-spin exchange interactions is attracting
much attention.  It has been found to be important in several
materials such as two-leg ladder compound
La$_x$Ca$_{14-x}$Cu$_{24}$O$_{41}$\cite{Brehmer,Matsuda,Schmidt},
two-dimensional antiferromagnet La$_2$CuO$_4$\cite{HondaKW,Coldea},
magnetism of two-dimensional quantum solids, e.g. solid ${}^3$He
films\cite{Roger}, and Wigner crystals\cite{Okamoto}.  
Four-spin ring exchange plays an essential role in several models
to give rise to
exotic phases due to its frustration, {\it e.g.},  the vector chirality phases\cite{PRL.95.137206},
nematic orderings\cite{PRL.96.027213}, and octapolar order\cite{PRL.97.257204}.

Especially,
the two-leg ladder model with the multiple-spin exchange interactions
has been studied extensively
\cite{HikiharaDualPRL,JPSJ.77.014709,MomoiDualPRB,Laeuchli,FourspinEntangle,GritsevFourspin,SatoFourspin}.
To clarify its rich phases,
not only correlation functions corresponding to phases
but also
entanglement concurrence\cite{PRB.74.155119} and string
order\cite{Laeuchli} are useful
to characterize the phases.
As such a novel order parameter,
which is beyond the Ginzburg-Landau symmetry-breaking description,
there is an order parameter based on the
topological invariants
corresponding to the topological
order\cite{TOrder}.

Recently, Berry phases\cite{Berryphase} 
have been used in order to detect the topological order
and the quantum order\cite{HatsugaiOrder1,HatsugaiOrder2,HatsugaiOrder3}.
The Berry phases are quantum quantities based on the Berry
connection which is defined by the overlap between the two states with
infinitesimal difference
and do not have
any corresponding classical analogues.
Then, one can define it even though there is no classical order parameter.
The advantage of the Berry phase is that it quantizes to $0$
or $\pi$ even in the finite sized systems in any dimension
when the system has the time reversal invariance. 
It has been
successfully applied to several quantum systems such as 
generalized valence bond solid states, dimerized Heisenberg models,
\cite{HatsugaiOrderP,VBSBerry,HatsugaiOrder3}, 
and $t$-$J$ model\cite{Maruyama}.
For these systems, the non-trivial ($\pi$) Berry phase on a link reveals a singlet on the link,
which is a pure quantum object due to two-spin exchange.

In this paper, we extend the quantized Berry phase to be sensitive to the 
effect of four-spin ring exchange interaction,
and apply it
to
a $S=1/2$ spin ladder with ring exchange interactions.
According to the phase diagram\cite{Laeuchli},
there are two first order transitions from ferromagnetic phase.
One is transition to the {\it rung singlet} phase.
The other is that to the {\it dominant collinear spin} phase,
which connect to the {\it dominant vector chirality} phase through crossover at a self-dual point\cite{JPSJ.77.014709}.
These two phases are the singlet phases with  short-range order and have dominant correlation of collinear spin and vector chirality respectively.
The rung singlet phase includes the spin ladder with only two-spin exchange interactions
and the dominant vector chirality phase includes that with only four-spin exchange interactions.
These phases have a unique ground state with finite gap,
while the other phases including ferromagnetic phase 
do not have finite gap under the translational symmetry.
Moreover, through the spin-chirality duality transformation,
the gap in the dominant vector chirality phase of the original Hamiltonian is smoothly connected to that in the rung singlet phase of the transformed Hamiltonian
\cite{HikiharaDualPRL}.
Although it is believed that the ground state of the rung singlet phase is well approximated by the product of local rung singlets,
there is no simple picture for the dominant vector chirality phase.
To clarify it,
we discuss the adiabatic connection of the Hamiltonian to a simple
model, i.e., a topologically equivalent model.

\section{Definition of the Berry phase}
Let us start with the definition of Berry phase\cite{Berryphase} in a quantum spin system.
For the parameter dependent Hamiltonian $H(\phi)$, the
Berry phase $\gamma$ of the ground state is 
defined as $i\gamma=\int_0^{2\pi}A(\phi)d\phi$ (mod $2\pi$), where $A(\phi)$ is the
Abelian Berry connection obtained by the single-valued
normalized ground state $|\mbox{gs}(\phi)\rangle$ of $H(\phi)$ as 
$A(\phi)=\langle\mbox{gs}(\phi)|\partial_\phi|\mbox{gs}(\phi)\rangle$.
This Berry phase is quantized to $0$ or $\pi$
if $|\mbox{gs}(\phi)\rangle$ is a gapped ground state
and the Hamiltonian $H(\phi)$ is invariant
under the anti-unitary operation $\Theta$, {\it i.e.} $[H(\phi),\Theta]=0$.
It has a remarkable property that the Berry phase has topological
robustness against the small perturbations unless the
energy gap between the ground state and the first excited state closes.
We note that the Berry phase is undefined 
if the energy gap vanishes while varying the parameter $\phi$.
Then, we limit ourselves to the rung singlet phase and
the dominant vector chirality (collinear spin) phase,
which have finite gap.
To calculate the Berry phase numerically\cite{PRB.47.1651}, 
we use $\gamma=\lim_{M\rightarrow\infty}\gamma_M$,
where $\gamma_M$ is defined  by discretizing the parameter space of $\phi$
into $M$ points as 
\begin{math}
\gamma_M=-\sum_{m=1}^{M}\mbox{arg}C(\phi_m)\mbox{,}\ \phi_m={2\pi m}/M,
\end{math}
where $C(\phi_m)$ is defined by
$C(\phi_m)=\langle \mbox{gs}(\phi_m)|\mbox{gs}(\phi_{m+1})\rangle$
with
$\phi_{M+1}=\phi_1$.
In general, $M$ can be a very small number\cite{Fukui}.

To study a local structure of the quantum system,
such as a local singlet,
we use a local spin twist on a specified link $i,j$
as the parameter $\phi$\cite{HatsugaiOrder3}.
Under this local spin twist on a link $i,j$, the term 
$S_i^{+}S_j^{-}+S_i^{-}S_j^{+}$
in the Hamiltonian is replaced with
$e^{i\phi}S_i^{+}S_j^{-}+e^{-i\phi}S_i^{-}S_j^{+}$, 
where $S^{\pm}_i=S_i^x\pm iS_i^y$.
Although previous studies\cite{HatsugaiOrder1,HatsugaiOrder2,HatsugaiOrder3,Maruyama,VBSBerry}
deal with the spin twist only for the two-body terms,
we extend the twist to the four-spin exchange interactions.
Since the Hamiltonian is written by $H=\sum_{ij} J_{ij} \bm{S}_i\cdot\bm{S}_j
+ \sum_{ijkl} K_{ijkl} (\bm{S}_i\cdot\bm{S}_j) (\bm{S}_k\cdot\bm{S}_l)$,
the local spin twist for a selected link $i,j$
is introduced into all the term $\bm{S}_i\cdot\bm{S}_j$ in the Hamiltonian.
As described below,
the extended Berry phase can detect not only the local singlet but also the plaquette singlet.

The quantized Berry phase is considered as a link-variable.
Then each link is labeled as one of three labels: ``$0$-bond'', ``$\pi$-bond'', or
``undefined''. 
Especially,
we shall calculate the leg Berry phase $\gamma_l$,
the rung Berry phase $\gamma_r$, and
the diagonal Berry phase $\gamma_d$.
The quantization of the Berry phase
is guaranteed
by the time reversal symmetry $\Theta$
of the quantum spin system\cite{HatsugaiOrder1}.

\section{The models and the results}
\subsection{$S=1/2$ spin ladder model with four-spin exchange
interaction}
\begin{figure}
 \includegraphics[width=9cm,height=6cm]{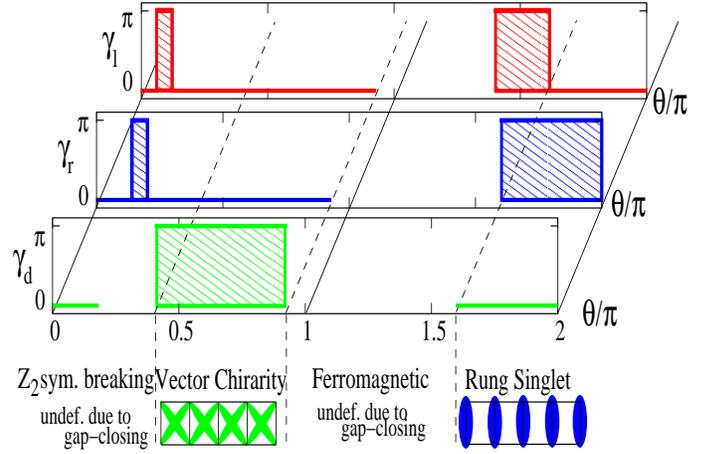}
\caption{(color online)
Leg, rung, and diagonal Berry phase $\gamma_l$, $\gamma_r$, $\gamma_d$  of the $S=1/2$ periodic $N=16$ 
ladder model with four-spin interactions
as a function of $\theta$ ($J=\cos\theta, K=\sin \theta$)
with the schematic pictures of corresponding phases.
Berry phases are zero,
$\pi$ (shaded region),
or
undefined (region without data).
}
\label{fig:Berryfour}
\end{figure}
The $S=1/2$ spin ladder model with four-spin exchange
interaction is described by the following Hamiltonian
\begin{eqnarray}
H_{\rm cyc} &=&J\left[
\sum_{i=1}^{N/2}\sum_{\alpha=1,2}
\bm{S}_{i,\alpha}\cdot\bm{S}_{i+1,\alpha}+
\sum_{i=1}^{N/2}\bm{S}_{i,1}\cdot\bm{S}_{i,2}
\right]
\nonumber\\
&+&K\sum_{i=1}^{N/2}(P_i+P_i^{-1}),
\end{eqnarray}
with the ring exchange
\begin{eqnarray}
  P_i + P_i^{-1}
  &=& \bm{S}_{i,1}\cdot\bm{S}_{i,2} + \bm{S}_{i+1,1}\cdot\bm{S}_{i+1,2}
+ \bm{S}_{i,1}\cdot\bm{S}_{i+1,1} \nonumber\\
&+& \bm{S}_{i,2}\cdot\bm{S}_{i+1,2} + \bm{S}_{i,1}\cdot \bm{S}_{i+1,2} + \bm{S}_{i,2}\cdot\bm{S}_{i+1,1}
\nonumber\\
&+& 4(\bm{S}_{i,1} \cdot \bm{S}_{i,2})
     (\bm{S}_{i+1,1} \cdot \bm{S}_{i+1,2}) \nonumber \\
&+& 4(\bm{S}_{i,1} \cdot \bm{S}_{i+1,1})
     (\bm{S}_{i,2} \cdot \bm{S}_{i+1,2})  \nonumber \\
&-& 4(\bm{S}_{i,1} \cdot \bm{S}_{i+1,2}) (\bm{S}_{i,2} \cdot
\bm{S}_{1,i+1}), \label{eq:P4}
\end{eqnarray}
where $\bm{S}_{i,\alpha}$ are the spin-$1/2$ operators on the site
$(i,\alpha)$ and $N$ is the total number of sites.
The periodic boundary condition is imposed as
$\bm{S}_{N/2+i,\alpha}=\bm{S}_{i,\alpha}$
for all of the models in this paper.
We set the parameters as $J=\cos{\theta}$, $K=\sin{\theta}$.
Figure \ref{fig:Berryfour} shows Berry phases on local links of $N$=16 ladder
obtained numerically by the exact diagonalization method.
Three kinds of Berry phases, 
$\gamma_l$ on the leg link $(i,1)-(i,2)$,
$\gamma_r$ on the rung link $(i,1)-(i,2)$,
and 
$\gamma_d$ on the rung link $(i,1)-(i+1,2)$,
are calculated
and showed translational symmetry.
Nontrivial ($\pi$) Berry phases obtained for finite size systems identified
two different phases;
One is the rung singlet phase and the other is the dominant vector chirality phase including the dominant collinear spin phase.
These three phases have a spin gap in thermodynamic limit\cite{Laeuchli,HikiharaDualPRL}.
The crossover between the dominant vector chirality phase and the dominant collinear spin phase is not found.
It is neither found by the entanglement\cite{PRB.74.155119} nor by the Lieb-Schulz-Mattis twist operator\cite{Laeuchli}.

Berry phases in the other phases are also obtained in Fig.~\ref{fig:Berryfour}
but becomes undefined due to gap-closing in the thermodynamic limit,
such as in the ferromagnetic phase
and around the self-dual point $\theta=\mbox{arctan}(1/2)\sim 0.14\pi$.
The staggered dimer and scalar-chirality phases around the self-dual point
are Z$_2$ symmetry breaking phase with two-fold degeneracy
in the thermodynamic limit\cite{PRB.74.155119}.
In a $N=16$ system,
there is a finite-size gap in the $S_z=0$ sector of $H_{\rm cyc}$
for most values of $\theta$
and we can obtain the Berry phase.
However,
results around the self-dual point in Fig.\ref{fig:Berryfour}
shows the numerical instability,
i.e., Berry phase becomes undefined.
Note that
we can define non-Abelian Berry phase to avoid the numerical instability
with using the gap above the two-fold degenerated ground states in the thermodynamic limit
around the self-dual point.
The transition point $\theta_c$ between the rung singlet phase and
Z$_2$ symmetry breaking phase
is about $0.1\pi$\cite{Laeuchli,HikiharaDualPRL},
while Fig.~\ref{fig:Berryfour} shows $\theta_c < 2\pi$.
To clarify this difference,
we shall study the Berry phase at $\theta=0$.

Before studying the model at $\theta=0$,
let us now interpret the phases by the
adiabatic modification 
in order to obtain a decoupled model with the same Berry phases.
We shall calculate the $\phi$ dependence of the energy gap
to see whether the gap closes or not during the adiabatic modification
since the Berry phase remains the same if the gap does not close during the
adiabatic modification.
\paragraph{Rung singlet phase}
We consider the adiabatic modification from $H_{\rm cyc}|_{\theta=1.8\pi}$ in the rung singlet phase to 
the completely decoupled model which has the Heisenberg type coupling only on
the rung bonds:
\begin{eqnarray}
H_{\rm RS}&=&\sum_{i=1}^{N/2}\bm{S}_{i,1}\cdot\bm{S}_{i,2}.
\end{eqnarray}
These two models are connected by adiabatic parameter $\alpha$ as
$H(\alpha) = (1-\alpha)H_{\rm cyc}+\alpha H_{\rm RS}$.
\begin{figure}[!tb]
 \includegraphics[width=7.5cm]{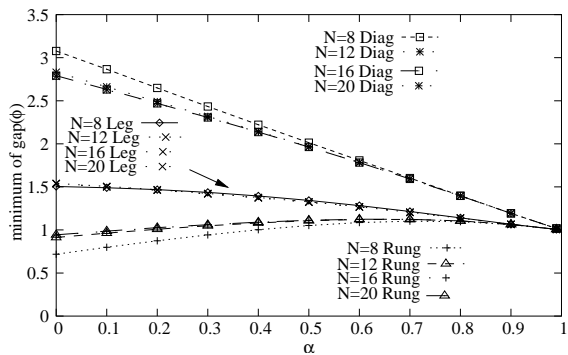}
\caption{
Minimum gaps for twist $\phi$ on a leg, rung, and diagonal link
of the $S=1/2$ periodic ladder model with four-spin interactions 
as a 
function of adiabatic parameter $\alpha$
for $H(\alpha) = (1-\alpha)H_{\rm cyc}\mid_{\theta=1.8\pi}+\alpha H_{\rm RS}$.
The several data are plotted for $N=8,12,16$, and 20.
}
\label{fig:RSGapfour}
\end{figure}
As shown in Fig.\ref{fig:RSGapfour}, 
the minimum gap through the twist $\phi$
does not close.
It means not only that the gap of $H_{\rm cyc}|_{\theta=1.8\pi}$
adiabatically connects to the singlet-triplet gap
but also that these two models with the same Berry phases are topologically identical.
This adiabatic connection is consistent with the fact that 
the ground state is well approximated by the product of local rung singlets.

\paragraph{Vector chirality phase}
We also consider the adiabatic modification from $H_{\rm cyc}|_{\theta=0.8\pi}$ in the dominant vector-chirality phase
to a decoupled model:
\begin{eqnarray}
H_{\rm DVC} &=& \sum_{i=1}^{N/4} \left(\bm{S}_{2i,1}\times\bm{S}_{2i,2}\right)\cdot
\left(\bm{S}_{2i+1,1}\times\bm{S}_{2i+1,2}\right).
\end{eqnarray}
We call its ground state as the ``dimerized
vector-chiral state'' since it minimize the local operator
$\left(\bm{S}_{2i,1}\times\bm{S}_{2i,2}\right)\cdot\left(\bm{S}_{2i+1,1}\times\bm{S}_{2i+1,2}\right)
=
\left(\bm{S}_{2i,1}\cdot \bm{S}_{2i+1,1}\right)\left(\bm{S}_{2i,2}\cdot\bm{S}_{2i+1,2}\right)
-
\left(\bm{S}_{2i,1}\cdot \bm{S}_{2i+1,2}\right)\left(\bm{S}_{2i,1}\cdot\bm{S}_{2i+1,2}\right)
$.
This operator is used as an order parameter in previous
studies
and a classical spin configuration is corresponds to a 90$^\circ$ spin structure\cite{GritsevFourspin}.
The ground state is the product of plaquette singlet states\cite{Todo}.
Moreover,
through the duality transformation\cite{HikiharaDualPRL},
$H_{\rm DVC}$ is mapped to
the summation of
\begin{math}
\tilde{\bm{S}}_{2i,1}\cdot \tilde{\bm{S}}_{2i,2}
+\tilde{\bm{S}}_{2i+1,1}\cdot \tilde{\bm{S}}_{2i+1,2}
- \tilde{\bm{S}}_{2i,1}\cdot \tilde{\bm{S}}_{2i+1,2}
- \tilde{\bm{S}}_{2i,2}\cdot \tilde{\bm{S}}_{2i+1,1}
.
\end{math}
This transformed model is identified as the rung singlet phase by the Berry phase.
It should be emphasized again that
the gap with no twist $\phi=0$ in the dominant vector chirality phase
is smoothly connected that in the rung singlet phase of the transformed Hamiltonian\cite{HikiharaDualPRL}.

\begin{figure}[!tb]
 \includegraphics[width=7.5cm]{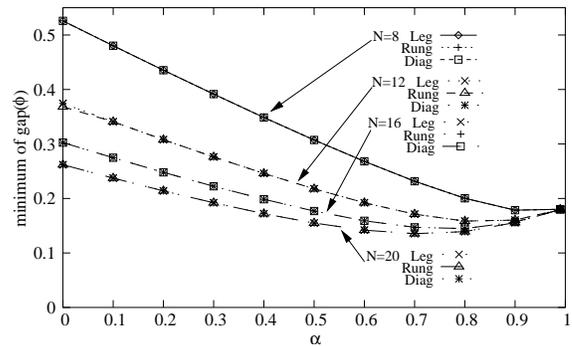}
\caption{
Minimum gaps for twist $\phi$ on a leg, rung, and diagonal link
of the $S=1/2$ periodic ladder model with four-spin interactions as a 
function of adiabatic parameter $\alpha$
for $H(\alpha) = (1-\alpha)H_{\rm cyc}\mid_{\theta=0.8\pi}+\alpha H_{\rm DVC}$.
The several data are plotted for $N=8,12,16$, and 20.
}
\label{fig:DPSGapfour}
\end{figure}

Figure \ref{fig:DPSGapfour} shows the minimum energy gap through the twist $\phi$
of $H(\alpha)=(1-\alpha) H_{\rm cyc}\mid_{\theta=0.8\pi} +  \alpha H_{\rm DVC}$
in the adiabatic deformation.
Since the gap does not close during the modification in $N=20$ system at least, 
we identify that the nontrivial diagonal Berry phase $\gamma_d$ exhibit
the decoupled vector-chiral state.
We should note that the modified Hamiltonian breaks the translational
symmetry although the original Hamiltonian does not.
In other words, although the gap does not close by the twist of diagonal bond 
in {\it odd}-th plaquette, it does for the diagonal bond in {\it even}-th
plaquette.
In this sense, the ground state of $H_{\rm cyc}|_{\theta=0.8\pi}$
is {\it locally} identical to the vector chiral state.

\subsection{$S=1/2$ spin ladder model without four-spin exchange
interaction}
Although
the rung singlet phase is thought to include
$\theta=0$\cite{Laeuchli},
Fig.~\ref{fig:Berryfour} shows that the rung singlet phase 
does not include $H_{\rm cyc}|_{\theta=0}$ ($J=1, K=0$).
To clarify it,
we study $S=1/2$ spin ladder model {\it without} four-spin exchange
interaction:
\begin{eqnarray}
H_{\rm lad}&=&J_{\rm l}\sum_{i=1}^{N/2}\sum_{\alpha=1,2}
\bm{S}_{i,\alpha}\cdot\bm{S}_{i+1,\alpha}
+J_{\rm r}\sum_{i=1}^{N/2}\bm{S}_{i,1}\cdot\bm{S}_{i,2}
,
\end{eqnarray}
where $J_{\rm l}$ and $J_{\rm r}$ are parametrized as $J_{\rm l}=\sin{\theta}$ and
$J_{\rm r}=\cos{\theta}$, respectively.
We consider the antiferromagnetic case of $0\leq \theta<\pi/2$ in this paper
to concentrate on $H_{\rm lad}|_{\theta=\pi/4} = H_{\rm cyc}|_{\theta=0}$.
Note that $H_{\rm lad}|_{\theta=0} = H_{\rm RS}$.
On the other hand,
$H_{\rm lad}|_{\theta=\pi/2}$ is the model of decoupled two chains,
which has gapless excitation in the thermodynamic limit.
\begin{figure}[!tb]
\includegraphics[width=9cm,height=6cm]{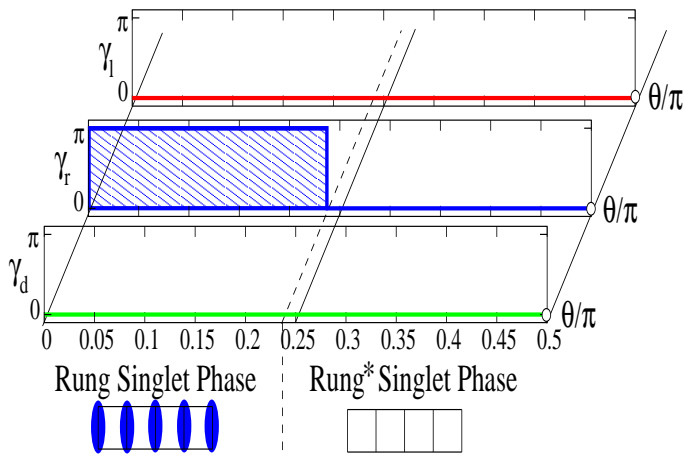}
\caption{(color online)
Leg, rung, and diagonal Berry phase $\gamma_l$, $\gamma_r$, $\gamma_d$  of 
the $S=1/2$ periodic $N=16$ 
ladder model without four-spin interactions
as a function of $\theta$ ($J_{\rm r}=\cos\theta, J_{\rm l}=\sin \theta$)
with the schematic pictures of corresponding phases.
Berry phases are zero,
$\pi$ (shaded region),
or
undefined (region without data).
}
\label{fig:Berrylad}
\end{figure}

Figure \ref{fig:Berrylad} shows the $\theta$ dependence of the
Berry phases on local links of $H_{\rm lad}$ at $N=16$
except for the gapless point $\theta=\pi/2$.
$\gamma_{d}=0$ is trivial because there is no diagonal interaction.
Although 
the gap of untwisted Hamiltonian $\phi=0$
is smoothly connected\cite{PRB.47.3196},
the gap of twisted Hamiltonian $\phi=\pi$ for $\gamma_r$ 
closes at $\theta\sim 0.233 \pi< \pi/4$
and
$\gamma_{r}$ changes.
We denote the $\gamma_{r}=0$ phase for $\theta>\pi/4$ as rung$^*$ phase,
which implies the limitation of the localized rung-singlet picture
and encourages us to use another picture such as the resonating valence bond theory\cite{PRL.73.886}.
It is consistent with 
very strong quantum fluctuations introduced on the product singlet ground state\cite{PRL.81.1941}
and
with ``non-perturbative'' (in $J_r/J_l$) behavior obtained numerically\cite{PRB.47.3196}.
To clarify the phase which has no $\pi$-bond,
we need further discussion 
with considering another kind of spin twist,
because
the twist should be corresponds to a local structure of the model.
It should be emphasized that
the rung$^*$ phase results from
the quantum phase transition of the twisted Hamiltonian ($\phi=\pi$) for $\gamma_r$
and is not contradicting previous studies for untwisted Hamiltonian.

\section{Conclusion}
In conclusion, we have shown that the quantized Berry phases is useful to classify the
phases of spin chains with four-spin interaction.  
In the dominant vector-chirality phase which comes from the ring exchange interaction,
the Berry phase on each diagonal link is used as a plaquette variable
and becomes a non-trivial value ($\pi$),
while the Berry phase has been used as a link variable in previous study.
The Berry phase is
also useful to clarify the phase boundary from the finite-size
systems since it is quantized even in the finite size systems.  The
property of the phase is revealed through the adiabatic deformation
into a decoupled model: the rung singlet phase (and the vector
chirality phase) is corresponds to a product of rung singlets
(plaquette singlets).
These two phases are connected through the duality transformation.
The other phases of this model
will be detected by the Berry phase with another kind of twist
or the non-Abelian Berry phase for two-fold degenerated ground states.

\begin{acknowledgments}
A part of the numerical diagonalization has been accomplished by utilizing 
the program package TITPACK ver. 2. The computation in this work has been
 done using the facilities of the Supercomputer Center, Institute for
 Solid State Physics, University of Tokyo.
This work has been supported
in part by Grants-in-Aid for Scientific Research,
No.20740214 from JSPS for IM,
No. 20340098, 20654034 from JSPS and 
No. 220029004, 20046002 on Priority Areas from MEXT.
\end{acknowledgments}

\bibliography{manuscript,../macro,../wiki}

\end{document}